# Pulse shape discrimination based on the Tempotron: a powerful classifier on GPU

Haoran Liu, Peng Li, Mingzhe Liu, Kaimin Wang, Zhuo Zuo, and Bingqi Liu

*Abstract*— This study utilized the Tempotron, a robust classifier based on a third-generation neural network model, for pulse shape discrimination. By eliminating the need for manual feature extraction, the Tempotron model can process pulse signals directly, generating discrimination results based on prior knowledge. The study performed experiments using GPU acceleration, resulting in over 500 times faster compared to the CPU-based model, and investigated the impact of noise augmentation on the Tempotron performance. Experimental results substantiated that Tempotron serves as a formidable classifier, adept at accomplishing high discrimination accuracy on both AmBe and time-of-flight PuBe datasets. Furthermore, analyzing the neural activity of Tempotron during training shed light on its learning characteristics and aided in selecting its hyperparameters. Moreover, the study addressed the constraints and potential avenues for future development in utilizing the Tempotron for pulse shape discrimination. The dataset used in this study and the GPU-based Tempotron are publicly available on GitHub at https://github.com/HaoranLiu507/TempotronGPU.

## I. Introduction

PULSE shape discrimination (PSD) is a radiation detection technique that enables the differentiation of various types of particles based on their unique pulse shapes [1]. Radiation detectors are susceptible to background radiation originating from cosmic, terrestrial, or artificial sources. Therefore, in many cases, a detection system must distinguish between particle types to facilitate accurate measurement and analysis. Radiation generates a distinctive pulse signal when it interacts with a detector capable of PSD. By examining the shape of the pulse, it becomes feasible to identify the specific type of particle that caused the detected radiation [2]. This identification is achieved through analyzing the difference in the triplet state induced by the incident particles, which subsequently leads to variations in delayed fluorescence and phosphorescence [3]. For instance, PSD can distinguish between neutron and gamma-ray radiation [4], which is important for applications such as nuclear reactors [5, 6], radiopharmaceuticals [7], and homeland security [8].

This technique relies on the fact that different types of radiation have different energy deposition patterns in detectors, which leads to distinct pulse shapes. In the example of neutron and gamma-ray discrimination, the pulse signals of these two particles have a similar rising edge but differ significantly in the rest of the pulse. The neutron is more likely to induce a distinctive interaction effect called inter-system crossing with the π-electron state of the scintillator molecule due to its greater mass. This effect leads to delayed fluorescence and phosphorescence ($10^{-4}$ s) that exhibit longer decay time compared to fluorescence ($10^{-9}$ s). As a result, the falling edge of a gamma-ray pulse is steeper than that of a neutron, as scintillation decays basically through fluorescence when induced by gamma-rays; hence, it is more rapidly when a gamma-ray photon interacts with the scintillator. Various PSD methods employ these distinctions in pulse signals to differentiate particle types, thereby improving the accuracy and reliability of radiation detection systems [4].

PSD can be achieved using various methods, which can be categorized into two classes: statistical discrimination and prior-knowledge discrimination. The statistical discrimination approach requires collecting a large number of radiation pulse signals, calculating a discrimination factor for each pulse, constructing a histogram for all discrimination factors, and differentiating particle types based on the statistical distributions observed in the histogram. Generally, the histogram of discrimination factors exhibits several Gaussian distributions corresponding to the number of particle types present in the dataset. The calculation of the discrimination factor from a radiation pulse signal is essentially a feature extraction process. This can be accomplished through various PSD methodologies such as traditional time-domain approaches (e.g., charge comparison [9], zero crossing [10, 11], and charge integration [12, 13]), frequency-domain methods

Manuscript received xx xxxxx 202x; revised xx xxxxx 202x; accepted xx xxxxx 202x. Date of publication xx xxxxx 202x; date of current version xx xxxxx 202x. This work was supported in part by the National Natural Science Foundation of China under Grant U19A2086, Grant 12205078, and Grant 42104174; and in part by the Wenzhou Major Science and Technology Innovation Project under Grant ZG2023011. (Corresponding author: Mingzhe Liu)

Haoran Liu, Mingzhe Liu, and Kaimin Wang are with the State Key Laboratory of Geohazard Prevention and Geoenvironment Protection, Chengdu University of Technology, Chengdu 610059, China, and also with the School of Data Science and Artificial Intelligence, Wenzhou University of Technology, Wenzhou 325000, China (e-mail: liuhaoran@cdut.edu.cn, liumz@cdut.edu.cn, wangkaimin@stu.cdut.edu.cn).

Peng Li and Zhuo Zuo are with the Engineering & Technical College of Chengdu University of Technology, Leshan 614000, China, and also with the Southwestern Institute of Physics, Chengdu 610225, China (e-mail: lipeng@stu.cdut.edu.cn, zuozhuo@stu.cdut.edu.cn).

Bingqi Liu is with the School of Mechanical Engineering, Chengdu University, Chengdu 610106, China (e-mail: liubingqi@cdu.edu.cn).

Color versions of one or more of the figures in this paper are available online at http://ieeexplore.ieee.org.

Digital Object Identifier

(e.g., fractal spectrum [14] and frequency gradient analysis [15]), and recently developed intelligent methodologies (e.g., quantum clustering [16], pulse-coupled neural networks [17], and ladder gradient [18]). These approaches have been well researched and developed in the PSD field and have been validated as effective and robust. However, conducting discrimination based on a dataset with a large number of pulse signals is not convenient for real-time online signal processing, and its performance varies based on the dataset quality (e.g., when there are too many pile-up events or excessive background noise). Additionally, these approaches generally require numerous parameter settings for different detection conditions, which can be cumbersome and reliant on individual experience.

On the other hand, prior-knowledge discrimination realizes PSD differently. This method obtains prior knowledge from a pre-labeled dataset of radiation pulse signals, then directly applies this knowledge to incoming radiation signals. Consequently, this method can perform PSD on the basis of single radiation event or multiple events simultaneously for parallel computation. Currently, a limited number of machine learning-based PSD methods belong to this category [19, 20], including support vector machines [21] and K-nearest neighbors regression [22]. However, these machine learning-based PSD approaches generally require a combination of feature extraction processes such as continuous wavelet transform, nonnegative matrix factorization, and nonnegative tensor factorization [21, 23]. These feature extraction approaches are cumbersome and computationally expensive, hindering the development of prior-knowledge discrimination.

Considering these limitations, this study applies a powerful classifier to the PSD field, called the Tempotron. This classifier eliminates the need for feature extraction processes for pulse signals, as it can directly process each pulse based on acquired prior knowledge and generate a discrimination result for a radiation pulse.

The Tempotron is a type of third-generation neural network [24], adhering to the classification standard established by Maass [25]. Third-generation neural networks focus on spiking neural networks, which exploit spike timing for efficient information processing. These models emulate actual neuronal dynamics, enhance biological plausibility, and find applications in neuromorphic hardware and deep learning algorithms [26]. As a spiking neural network, the Tempotron employs an integrate-and-fire neuron model and utilizes gradient-based supervised learning for processing spike patterns containing spatiotemporal information. It serves as a potent binary classifier capable of effectively analyzing differences between pulse shapes. Readers unfamiliar with deep learning and machine learning fields please refer to [27, 28].

In this study, we examine the Tempotron learning behavior within the realm of PSD, assessing its performance and robustness. It is demonstrated that the Tempotron learns rapidly; even with a minimal number of epochs on a small training set, it can achieve relatively high discrimination accuracy. When trained for a more extended period, the Tempotron can attain exceptional discrimination accuracy, which remains highly robust and stable even when exposed to extreme levels of noise in the system.

Furthermore, this study implements the Tempotron using PyTorch, capitalizing on the formidable computational capability of the graphics processing unit (GPU). This accelerates its training, testing, and validation speeds by more than 500 times compared to the CPU-based Tempotron. The GPU-based Tempotron is made open-source and available on GitHub at https://github.com/HaoranLiu507/TempotronGPU. This open-sourced GPU-based Tempotron application is not limited to PSD but can be adapted to address various classification problems by adjusting the input signal, such as image and voice classification.

## II. METHODOLOGIES

### A. Tempotron

Tempotron is a spiking neural network model designed to decode and process complex spatiotemporal patterns in an efficient and effective manner [24]. Inspired by the incredible power of the human brain, the Tempotron employs neurons as fundamental computational components, facilitating memory and learning through the adjustment of synaptic efficacies. Moreover, the Tempotron conveys information using discrete spikes [29, 30]. By capitalizing on the unique temporal dynamics at its core, this model enhances the representation of real-world event sequences, thereby improving performance in various applications.

As depicted in Fig. 1a, the dendrites of the Tempotron (blue) establish several synapses with the terminal end of the presynaptic neuron axon (orange). In biological systems, synapses function as information transmission units between two neurons, modulating the contribution of incoming spikes to the internal activity of postsynaptic neuron by adjusting the quantity and type of neurotransmitters they release [31, 32]. Similarly, the Tempotron achieves this functionality by adjusting the efficacies that determine the connection strength of each synapse [33]. Upon receiving encoded spike patterns illustrated in Fig. 1b, the Tempotron exhibits corresponding internal activities within its soma, as depicted in Fig 1c. If this internal activity surpasses the threshold, a spike is generated, transmitted through the axon, and ultimately detected at the terminus of axon.

The Tempotron utilizes a leaky integrate-and-fire neuron model, with an exponentially decaying characteristic of its postsynaptic potentials (PSPs). The internal activity of the Tempotron $V(t)$, also known as the subthreshold membrane voltage, is influenced by a weighted sum of PSPs contributed by all synapses connected to the Tempotron neuron. Its mathematical expressions are defined as follows,

$$V(t) = \sum_i \omega_i \sum_{t_i} K(t - t_i) + V_{rest}, \qquad (1)$$

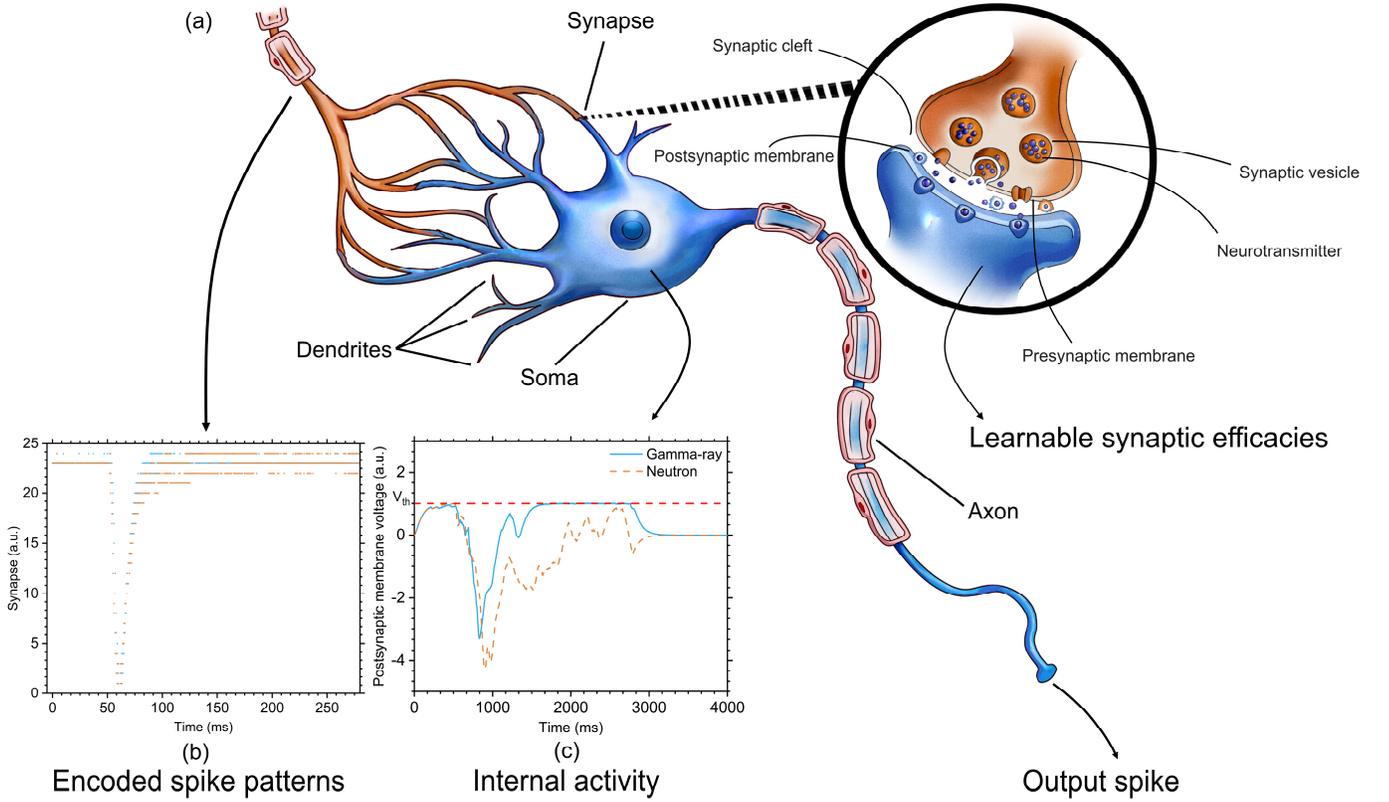

Fig. 1. Tempotron model. (a) Schematic representation of the Tempotron neuron. The dendrites of the Tempotron (blue) form multiple synapses with the terminal end of the presynaptic neuron axon (orange). The Tempotron receives (b) encoded spike patterns from the presynaptic neuron, subsequently exhibiting corresponding (c) internal activity within its soma. If this internal activity exceeds a threshold, a spike is generated, propagated through the axon, and ultimately detected at the terminus of axon.

$$K(t - t_i) = V_0 [e^{-\left(\frac{t-t_i}{\tau}\right)} - e^{-\left(\frac{t-t_i}{\tau_s}\right)}], \quad (2)$$

$$Y(t) = \begin{cases} 1, & V(t) > V_{th} \\ 0, & V(t) < V_{th} \end{cases}, \quad (3)$$

where, $t$ indicates the temporal index of the Tempotron neuron; $\omega_i$ represents the synaptic efficacy of the $i$ th synapse; $t_i$ describes the spike times of the $i$ th synapse; $K(t - t_i)$ represents the normalized PSP kernel that characterizes the effect of a spike; $V_{rest}$ represents the resting state membrane potential of the Tempotron neuron; $V_0$ is the normalization factor that ensures the uniform amplitude of postsynaptic potentials from distinct synapses, which is solely affected by the synaptic efficacy values; $\tau$ and $\tau_s$ indicate the time constants that control the exponential decay of the membrane integration and synaptic currents, respectively; $Y(t)$ indicates the output of the Tempotron neuron, with the value of 1 implying a spike generation and 0 indicating the absence of spike generation; $V_{th}$ is the threshold of the Tempotron, triggering a spike generation when the internal activity of a neuron surpasses its value.

The Tempotron neuron is a binary classifier that operates on a gradient-based supervised learning rule, with a spike and no-spike as the only two potential output states. The Tempotron learning process for this classification task is both uncomplicated and straightforward. Its two output states correspond to two kinds of input signals. In the present study, the classification task distinguishes between neutrons (0) and gamma-rays (1). The adjustment of the synaptic efficacies occurs during an output error, represented by the formula below,

$$\Delta \omega_i = \begin{cases} \lambda \sum_{t_i < t_{max}} K(t_{max} - t_i), & if\ Y = 0 \\ -\lambda \sum_{t_i < t_{max}} K(t_{max} - t_i), & if\ Y = 1 \end{cases}, \quad (4)$$

where, $t_{max}$ represents the moment at which $V(t)$ achieves its maximum value; $\lambda$ refers to the maximal synaptic update; and all other variables are identical to those in (1), (2), and (3).

*B. Pulse signal encoding*

As a third-generation neural network model, Tempotron receives and outputs information in the form of precise spike timings. This type of processing requires the encoding of input signals into spike trains. In this study, the encoding method involves using latency to encode time-domain pulse signals into spike patterns onto one axon terminal, and also involves the use of Gaussian receptive field (GRF) encoding to further encode spike patterns from a single axon onto multiple axon terminals in spatiotemporal patterns.

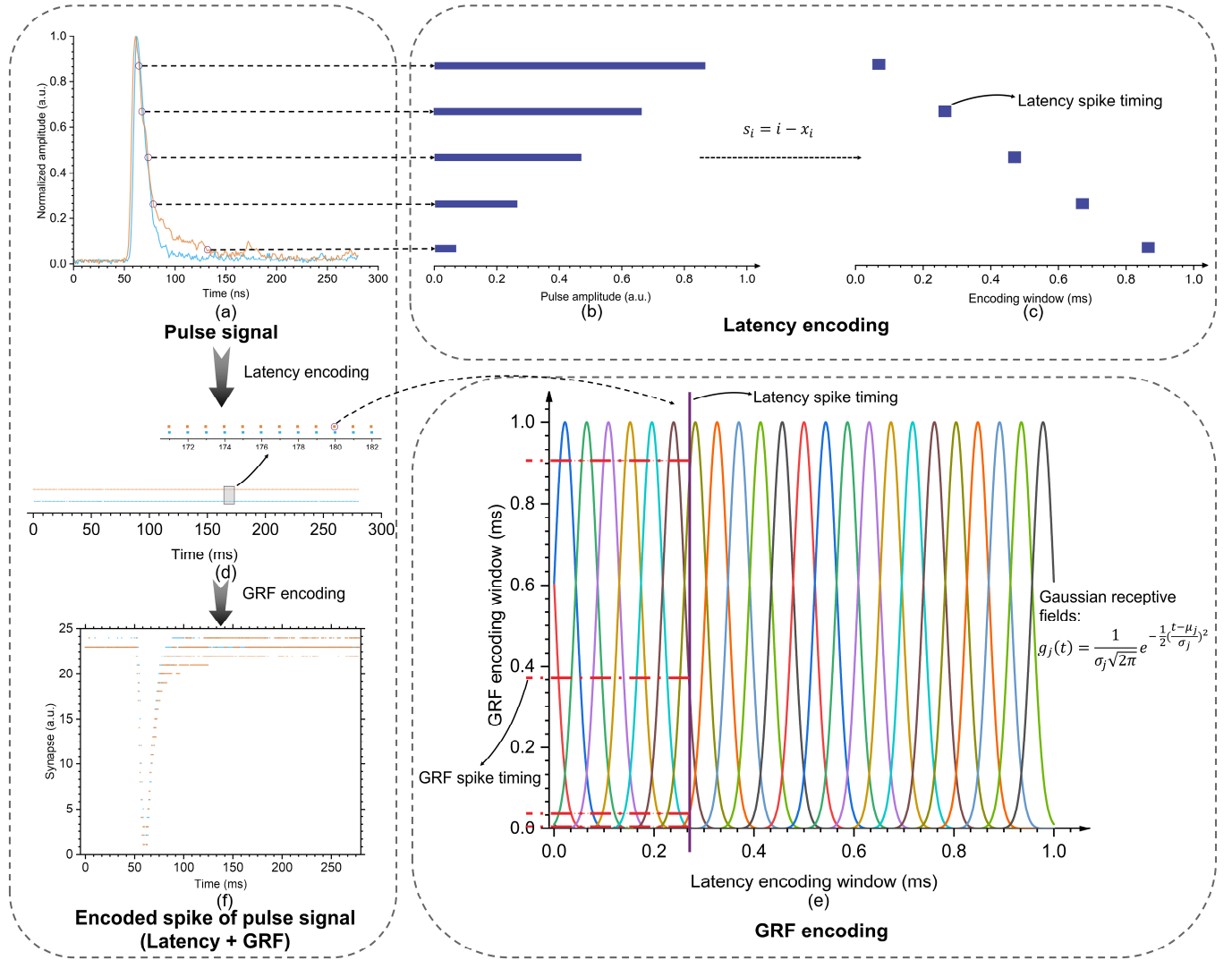

Fig. 2. Signal encoding scheme. Firstly, (a) normalized pulse signals are encoded using the latency method by assigning each sample point in a pulse signal to an encoding window. (b) The pulse amplitude is then mapped to the precise spike timing in (c) the encoding window. Next, the encoded time windows for all sample points in a pulse signal are merged together in an indexed order to form (d) the spike train for the signal. This spike train on a single presynaptic axon terminal is then encoded onto multiple terminals using (e) the Gaussian receptive field (GRF) encoding method. In this method, a spike timing in a latency encoding window interacts with multiple GRFs to determine the interaction location on the Y-axis of each GRF. Finally, the GRF spike timings of all GRFs' encoding windows are combined to produce (f) the final encoded spatiotemporal spike patterns.

Fig. 2 illustrates the signal encoding scheme. Neutron and gamma-ray pulse signal amplitudes are normalized to a 0-1 range, as shown in Fig. 2a. Each sample point $x_i$ in the pulse signal is assigned a 1 ms encoding window, with its amplitude used to calculate the spike time $s_i$ in the window according to the following formula,

$$s_i = i - x_i \quad (5)$$

where, $x_i$ denotes the $i$th sample position; $i$ denotes the $i$th encoding window; and $s_i$ denotes the precise spike timing in the $i$th encoding window.

As depicted in Fig. 2b and 2c, the closer the pulse amplitude is to 1, the earlier the spike time occurs. After all sample points are encoded in their respective encoding windows, the spike timings are composed based on their encoding window index, as illustrated in Fig. 2d. At this stage, pulse signals are transformed into spike patterns within one axon terminal. While the Tempotron can analyze information directly from these spike patterns, retrieving information from a single synapse proves to be inefficient and can impede the Tempotron learning process. A singe tunable synaptic efficacy will significantly restrict the learning capacity of the Tempotron neuron.

Consequently, the GRF encoding method is used to encode spike patterns onto multiple axon terminals, as shown in Fig. 2e. Each spike timing is re-encoded by placing multiple Gaussian curves in its encoding window, mathematically given as follows,

$$g_j(t) = \frac{1}{\sigma_j\sqrt{2\pi}} e^{-\frac{1}{2}(\frac{t-\mu_j}{\sigma_j})^2}, \quad (6)$$

where, $g_j$ denotes the $j$th GRF, a function of time $t$ in an encoding window; $\sigma_j$ is the standard deviation of the $j$th GRF, typically constant across all GRFs; and $\mu_j$ is the mean of the $j$th GRF, with the mean of all GRFs evenly distributed through the encoding window.

Each Gaussian curve represents an axon terminal, and a pulse intersects with multiple Gaussian curves. The Y-axis values of these intersections correspond to the spike timings of each axon terminal during the given time window. When the encoding of all time windows was completed, the pulse sequence from a single axon terminal was transferred to multiple axon terminals. This process enables the spatiotemporal pulse sequence to be transmitted to the neuron through multiple dendrites, providing the neuron with multiple synaptic efficacies to learn, and significantly enhancing its learning ability.

In summary, the initial radiation pulse signals (Fig. 2a) undergo the process of encoding into spike patterns, represented by one axon terminal (Fig. 2d) through latency encoding. Subsequently, these spike patterns are further transformed into spatiotemporal spikes, depicted on multiple axon terminals, using the GRF encoding method (Fig. 2f). Following the aforementioned encoding procedures, each pulse signal was encoded into spatiotemporal spike patterns distributed across several axon terminals of a presynaptic neuron, depicted as the orange neuron in Fig. 1. These axon terminals can form multiple synapses with the dendrites of a postsynaptic Tempotron neuron, thereby transmitting spatiotemporal spike patterns containing abundant information.

## III. EXPERIMENTS

### A. Experimental setups

The study utilized radiation pulse signals obtained from a $^{241}$Am$^9$Be neutron source to discriminate pulse shapes, which theoretically exhibits a 0.57 ± 10% gamma-ray to neutron ratio [34]. The dataset comprised of roughly 10,000 pulses. Detecting these pulse signals were a plastic EJ299-33 scintillator, and formed by a TPS2000B oscilloscope. Preprocessing removed corrupted signals resulting from events such as inadequate energy decomposition and pile-up, thereby excluding them from the final dataset. This preprocessing specifically targeted strongly corrupted pulses, including extended flat peaks and multiple peaks. No filtering or other denoising techniques were employed on the dataset. The study also utilized a time-of-flight (ToF) dataset measured from a $^{238}$Pu$^9$Be source using yttrium aluminum perovskite activated by cerium (YAP:Ce) gamma-ray detectors [35] and a NE213A fast neutron detector [36].

The pulse-coupled neural network (PCNN) method was used to label all signals due to its outstanding discrimination performance, as demonstrated in [17]. To select a training set from the discrimination results of PCNN, 500 gamma-ray pulses and 500 neutron pulses were chosen from the central portion of the Gaussian distribution. The distributions were generated by creating histograms of the discrimination factors for all signals. This set of 1,000 signals served as the training dataset for both the Tempotron and other machine learning approaches. Specifically, 80% of them were randomly assigned to the training set, and the remaining were assigned to the validation set. The remaining part of the dataset, comprising approximately 9,000 signals, was used as the testing set. The Tempotron was developed using PyTorch in Python, while other comparison methods were implemented using both MATLAB and Python. The experiments were conducted on an NVIDIA RTX 4090 GPU and an Intel i7-13700K CPU.

### B. Tempotron training

A subset of 1,000 pulse signals, representing around 10% of the entire dataset, is employed for Tempotron training. Amon them, 200 signals are reserved for validation, while 800 signals are used for training.

The learning rate of the Tempotron is provided in the form of an interval, for instance, $[10^{-6}, 10^{-3}]$. During training, the learning rate begins with the upper limit and is subsequently reduced by half after every twenty epochs. The mathematical expressions are given as follows,

$$lr = \begin{cases} lr_{up}, & if\ epoche = 1 \\ \frac{(lr-lr_{low})}{2}, & if\ epoche|20' \end{cases} \quad (7)$$

Furthermore, momentum learning is employed to accelerate the training of the Tempotron. For every synaptic efficacy adjustment, the efficacy changes $dw$, which are defined as the product of the learning rate $lr$ and $\Delta\omega_i$, are logged, and they modulate the $dw$ in the following adjustment, using the following formula:

$$dw_k = \varepsilon dw_{k-1} + (1-\varepsilon)dw_k, \quad (8)$$

where, $dw_k$ represents the efficacy changes at the $k$th synaptic efficacy adjustment and $\varepsilon$ denotes the momentum factor.

Moreover, noise augmentation is employed to improve the generality performance of the Tempotron. Three types of noise are added to the training dataset for this purpose. The first type involves adding Gaussian noise directly to the original signals with a factor, $\sigma_G$, which represents the sigma value of the Gaussian distribution. The second type entails adding jitter noise [37] that introduces random variation to the encoded spike times, and this variation also follows a Gaussian distribution, with a factor of $\sigma_j$ that reflects the sigma of the Gaussian distribution. The third type of noise, adding&missing [37], involves adding or deleting a spike randomly from the encoded spike patterns. It adjusts the spikes using a probability $p$.

Fig. 3 illustrates the training and testing loss of the Tempotron. It was trained under both $V_{th} = 1$ and $V_{th} = 10$ conditions. The Tempotron was trained with the following parameters: $\tau = 8.4$, $\tau_s = 2.1$, $dendrites\_num = 25$ (the number of presynaptic axon terminals that feeds information to the Tempotron), $epochs = 300$, $lr = [10^{-6}, 10^{-3}]$. The training employed both noise augmentation (with $\sigma_G = \sigma_j = p = 10^{-4}$) and momentum acceleration. The loss is calculated as the error rate between the results of Tempotron and the ground truth. The figure reveals that the Tempotron can reach a very low loss rate within roughly 100 epochs. Following the initial rapid decline, the loss finally converges to 0.05, which corresponds to a discrimination accuracy of approximately 95%. Notably, Tempotron with $V_{th} = 10$ converges even more rapidly than under the $V_{th} = 1$ condition, achieving 95% discrimination accuracy in only 50 epochs. This learning performance is a testament to the classification capabilities of Tempotron, as it can quickly learn to differentiate between two types of particles and perform PSD with high accuracy.

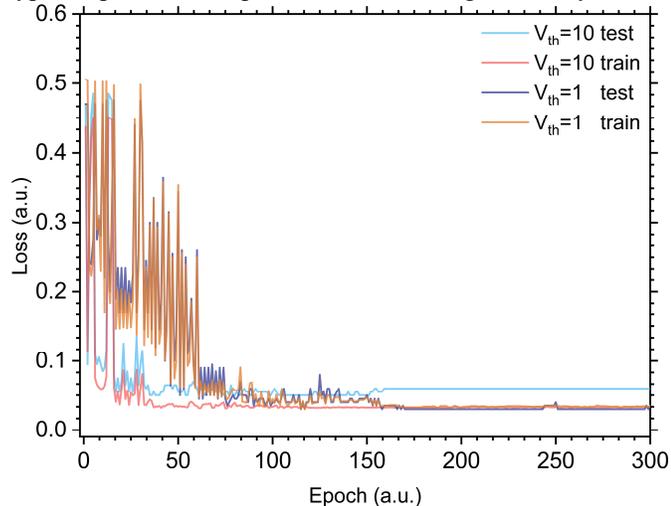

Fig. 3. Training performance of the Tempotron. The Tempotron loss converged very rapidly, reaching a stable low value within 100 epochs.

A comparison of discrimination performance among various methods was conducted to demonstrate the efficacy of the Tempotron. These methods consisted of statistical discrimination methods in the time-domain, frequency-domain, and intelligent algorithms, as well as pre-training-based machine learning methods. Charge comparison (CC) [9], zero crossing (ZC) [10, 11], charge integration (CI) [38-40], and pulse gradient analysis (PGA) [3] were the most commonly used time-domain statistical methods in the PSD field. Frequency gradient analysis (FGA) [15] was chosen for the frequency-domain method due to its robustness and ease of implementation. The intelligent methods included the recently proposed HQC-SCM [41], and ladder gradient (LG) [18] which exhibited remarkable discrimination performance. Finally, the machine learning methods consisted of a k-nearest neighbors (KNN) [22] regression approach and two simple multilayer perceptron (MLP) [42] models in the deep learning field. MLP-I was a single-layer perceptron with 280 neurons and a sigmoid activation function. In contrast, MLP-II was an eight-layer perceptron model with over 1,000 neurons overall, a rectified linear unit (ReLU) activation in each hidden layer, and a sigmoid activation in the output layer.

TABLE I
DISCRIMINATION ACCURACY OF VARIOUS METHODOLOGIES

| Method | CC | ZC | CI | PGA | FGA | PCNN |
|---|---|---|---|---|---|---|
| Accuracy | 0.9963 | 0.9193 | 0.9833 | 0.9813 | 0.9079 | - - |
| Method | HQC-SCM | LG | KNN | MLP-I | MLP-II | Tempotron |
| Accuracy | 0.9977 | 0.9774 | 0.9999 | 0.9982 | 0.9982 | 0.9534 |
| Trainable Parameters | - - | - - | - - | 301 | 329,345 | 25 |

As shown in Table I, for these data and experimental setup, all PSD methods achieved an accuracy of at least 90%. The pulse signals utilized in this experiment were actual measured signals, which renders the precise ground truth label of a signal unknown. Therefore, we selected the PCNN method as the ground truth, labeling all pulse signals and assembling the training datasets for methods that require training. Thus, the accuracy of PCNN is not provided in Table I. In the following ToF experiment, the accuracy of PCNN is proved to be highly consistent with the ground truth. Despite being proven efficient and robust, the discrimination results of the PCNN method cannot be interpreted as entirely correct, and some pulse signals may have been mislabeled. As a result, methods with accuracy above 90% offer acceptable discrimination performance, while those above 95% exhibit exceptional discrimination performance. The methods in the 95%-100% range of accuracy perform at the same level, with no higher or lower distinctions. Moreover, the accuracies are on the testing set.

Only the ZC and FGA methods exhibit a slightly higher accuracy than 90%. This finding corresponds to the commonly held belief that such fast-discrimination methods often perform worse than other discrimination methodologies. The accuracy of CC, PGA, HQC-SCM, and LG are close to 100%, implying that their discrimination outcomes are highly consistent with PCNN. Three distinctive discrimination methodologies producing similar results provide further evidence of the reliability of these methods and the ground truth labels assigned to the signals.

Unlike the statistical methods examined above, machine learning-based methods do not perform discrimination factor calculations to identify the Gaussian distribution of each particle type. Instead, these training-based methods draw prior knowledge solely from the ground truth label, which is based on the results produced by PCNN in this experiment. Therefore, they seek to achieve the same discrimination performance as PCNN, rather than uncovering pulse distinctions between particles in a fundamentally different manner. The experimental data demonstrated that both the KNN and two MLP methods reached an accuracy of almost 100%, which indicates that these methods successfully replicated the discrimination capability of PCNN. Note that the discrimination accuracies on the training, testing, and validation sets are consistent across all machine learning methods, suggesting the absence of overfitting. Moreover, it should be emphasized that the single-layer MLP-I also attained remarkably high accuracy, which is a typical linear classifier that cannot classify nonlinear datasets. Its strong performance suggests that the PSD problem after PCNN feature extraction is a linearly separable problem, particularly after pre-processing to eliminate corrupted pulses.

The Tempotron exhibits an accuracy of over 95%, indicating its successful replication of discrimination capability of PCNN. However, this accuracy is slightly lower than that of the other three machine-learning methods. We speculate that this difference in accuracy may stem from the fact that the Tempotron is a nonlinear classifier, which is susceptible to overfitting when dealing with linearly separable datasets, making it difficult to achieve accuracy close to 100%. Nevertheless, this does not imply that the performance of the Tempotron is inferior to that of other machine learning-based methods. The overfitting of the Tempotron is based on the assumption that the ground truth labeled by PCNN is linearly separable. As mentioned earlier, the labeling results of PCNN are not entirely correct. For some signals with unclear features caused by, for example, insufficient energy deposition or system noise, they are simply categorized by the linear classifier, resulting in certain discrimination errors. In contrast, the Tempotron must find a boundary around these signals that are challenging to classify based on their labeling, resulting in subtle differences between its classification results and those of the linear classifier. However, these differences may arise from either misclassifying a small number of signals that are correctly labeled or accurately classifying a small number of signals that are mislabeled, and overall, they do not have a significant impact on the effectiveness of PSD.

*C. Noise augmentation*

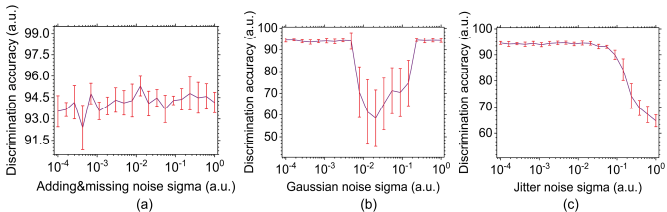

Fig. 4. Testing set accuracy under noise augmentation. (a) Adding&missing noise augmentation. (b) Gaussian noise augmentation. (c) Jitter noise augmentation.

Experiments were conducted to determine the noise level introduced by the noise augmentation process. The findings are presented in Fig 4. The Tempotron was trained with the following parameters: $\tau = 8.4$, $\tau_s = 2.1$, $dendrites\_num = 25$, $epochs = 300$, $lr = [10^{-6}, 10^{-3}]$. The training employed both noise augmentation (with $\sigma_G = \sigma_j = p = [10^{-4}, 1]$) and momentum acceleration. In the case of adding&missing noise, the Tempotron did not exhibit a drop in testing accuracy when the probability of introducing or deleting a spike ranged from 1 in 10,000 to 1. This result indicates that the Tempotron is highly resistant to noise in the presence of spikes. It is worth noting that when the probability of adding&missing noise approaches 1, nearly all the encoding windows with a spike will have the spike removed, and every encoding window without a spike will have a spike added. As a result, the information presented in the final augmented dataset only represents another encoding version of the original data rather than having a high level of random noise. As long as the number of dendrites is sufficient, the Tempotron can learn different spike patterns in a single neuron, and therefore demonstrate similar performance to the condition without noise augmentation.

The Tempotron shows stable performance for a wide range of noise levels in the case of Gaussian noise. However, its accuracy drops from 95% to approximately 60% when the sigma of Gaussian noise is around 0.01. The performance of the Tempotron then rises again as the noise sigma continues to increase until it reaches the same accuracy as the beginning when sigma is 1. These results, once again, demonstrate the high noise resistance of Tempotron. The performance drop observed during the middle-range noise levels is due to signal corruption, which significantly diminishes the difference between pulse signals of different particles. Nevertheless, when the noise level becomes too high to completely bury information inside a pulse signal, the noise-augmented dataset becomes a set of random signals with random labels. The Tempotron then treats this random dataset as background noise by attempting to minimize the loss on the training set as much as possible. Because there is still an original dataset directly from the encoded signal in the training set, the Tempotron finds the best weights to achieve the lowest loss on the non-augmented dataset, ignoring the random augmented dataset. These weights can make the loss on the testing and validation set converge, thereby achieving a similar result as the non-augmented condition.

In the case of jitter noise, the Tempotron showed stable performance across a wide range of noise levels. However, its accuracy significantly decreased as the noise level became too high. In contrast to the results observed with Gaussian noise, the high-noise scenario with jitter noise did not produce a similar outcome. Although the same sigma was used in both cases, the influence of jitter noise was much weaker than that of Gaussian noise. Consequently, a significantly larger sigma would be required to transform the jitter noise-augmented dataset into random signals with random labels.

In summary, the three noise augmentation methods are beneficial to the training of the Tempotron and can improve generality when appropriate noise intensity is used. However, when the noise intensity is too high, adding&missing noise loses its characteristic of random noise. Gaussian noise leads to a drop in accuracy initially and then becomes a random dataset that only slows down training speed. Similarly, jitter noise also leads to a drop in accuracy.

*D. Tempotron discrimination on GPU*

This study utilized PyTorch in Python to vectorized the computational process of signal encoding, training, testing, and validation.

The initial stage of the data processing involves encoding the data into spike patterns that can be readily processed by the Tempotron. In this experiment, the training dataset comprises 1,000 signals, each containing 280 sample points. Therefore, the dataset is transformed into a tensor with a shape of (1,000, 280) for encoding purposes. To ensure unified amplitude in the latency encoding approach, row normalization is carried out on the tensor. A latency encoding window is utilized, with a one-dimensional tensor of ascending numbers from 1 to 280 serving this purpose. The latency spike times are obtained by employing (5) to subtract the dataset tensor from the encoding window tensor. Notably, this one-dimensional tensor is automatically broadcast into the same shape as (1,000, 280) in Python, resulting in vectorized encoding of all signals' sample points. Subsequently, spike generation by weak amplitude is eliminated, with NaN replacing such sporadic spikes. The next step involves preparing a tensor of GRFs' means with a shape of (number of dendrites, 1,000, 280) to facilitate the formation of Gaussian receptive fields by (6). The number of postsynaptic dendrites corresponds to the number of presynaptic axon terminals because only dendrites involved in synapses are considered. Gaussian spike times are computed by applying the spike times obtained from the latency encoding as a variable to the GRF functions, with sigma being universal for all GRFs and the distribution of means from 0 to 1 being even. The Gaussian spike times for all signals' latency spike times on all dendrites are generated simultaneously in a vectorized manner by broadcasting the latency spike times and universal sigma into the shape of the tensor of GRFs' means. However, the intersection points at which the GRF function value is infinitely close to zero is meaningless. Thus, weak amplitude spikes are eliminated by thresholding Gaussian spike times, resulting in NaN. The encoding process concludes at this stage, with the dataset being made ready for Tempotron processing.

In the training process, the Tempotron processes input signals as a three-dimensional tensor with the shape (number of dendrites, batch size, sample points). The batch size represents the number of signals processed simultaneously. The postsynaptic currents caused by input spatiotemporal spike patterns are calculated by the Tempotron, according to (2). These currents are determined using the synaptic efficacies and the GRF spike times, obtained from the previous procedure. Notably, the internal neuronal activity is a weighted combination of all post-synaptic currents based on (1). Further, a logical statement following (3) is applied to determine whether an encoded input signal needs spike emission, and this emission corresponds to the signal label. If the above condition is satisfied, there is no need for synaptic efficacy modification. Conversely, for signals that require synaptic efficacy adjustments, the amount of adjustment is determined using (4). Next, the learning rate is multiplied by the adjustment amount before taking the average to obtain the present synaptic efficacies adjustment amount. The above-described steps are iteratively executed throughout the epochs until the Tempotron model has converged.

The testing and validation procedures used here are alike to the training process, except for the non-inclusion of synaptic efficacy adjustments. In summary, the Tempotron discrimination process on GPU exhibits computational efficiency utilizing vectorized calculation and GPU acceleration. It significantly accelerates the processing of numerous signals with multiple sample points, compared to the one-by-one calculation approach of CPU-based Tempotron.

*E. Neural activity of the Tempotron*

The synaptic efficacy adjustment properties are illustrated in Fig. 5. The Tempotron was trained with the following parameters: $\tau = 8.4$, $\tau_s = 2.1$, $dendrites_{num} = 20 / 80$, $epochs = 300$, $lr = [10^{-6}, 10^{-3}]$. The training employed momentum acceleration. Fig. 5 presents the adjustment amounts between epoch 51 and 52 of two Tempotron models (20 dendrites and 80 dendrites), as depicted in Fig. 5a and 5b, respectively. The adjustments are minor, with most synapses' efficacies changing at the $10^{-4}$ level for both Tempotrons. However, while the changes in synaptic efficacies are similar for both Tempotron models, their output discrimination accuracies differ significantly. Fig. 5c and 5d demonstrate an approximate 30% increase in the accuracy of the 20 dendrites Tempotron from epoch 51 to epoch 52, whereas the accuracy of the 80 dendrites Tempotron remains stable, as shown in Fig. 5e and 5f. This phenomenon arises due to the varying impact of minor synaptic efficacy changes on the internal activity of the Tempotron neuron. The influence is relatively subtle when the Tempotron can receive information from a large number of dendrites, whereas even minor changes in synaptic efficacies can make a significant difference when the number of dendrites is small. Consequently, the convergence course of the Tempotron is more volatile when the dendrite number is small, as presented in Fig. 3, making the training process easier but more unstable. On the other hand, when the number of dendrites is large, the convergence course is more stable, albeit at the cost of harder training.

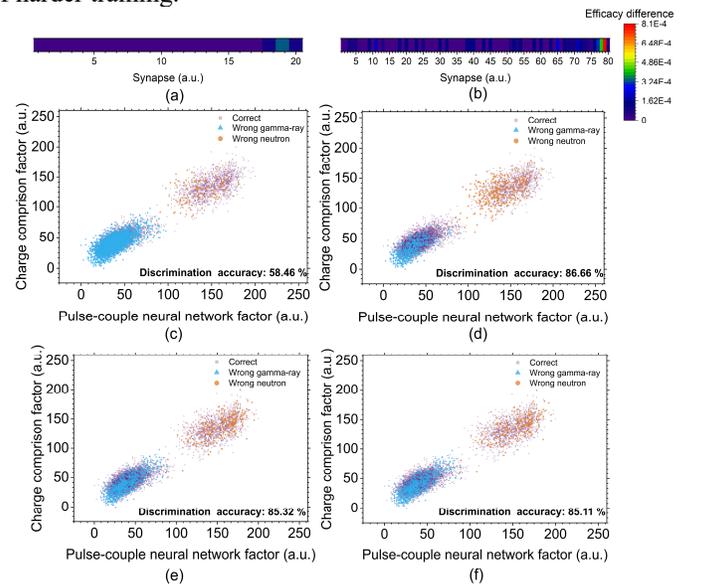

Fig. 5. Characteristics of synaptic efficacy adjustment. (a) Efficacy difference between epoch 51 and epoch 52 of the Tempotron model with 20 dendrites. (b) Efficacy difference between epoch 51 and 52 of the Tempotron model with 80 dendrites. The discrimination results and accuracies of (c) epoch 51 and (d) epoch 52 under the 20 dendrites condition, and of (e) epoch 51 and (f) epoch 52 under the 80 dendrites condition. The ground truth labels were given by a recognized PSD algorithm, specifically the PCNN.

Fig. 6 presents the convergence of synaptic efficacy during the training process. The Tempotron was trained with the following parameters: $\tau = 8.4$, $\tau_s = 2.1$, $dendrites_{num} = 25$, $epochs = 300$, $lr = [10^{-5}, 10^{-3}]$ for Fig. 6a and $lr = 10^{-3}$ for Fig. 6b and 6c. The training employed momentum acceleration. Fig. 6a depicts three independent training instances with random initialization of synaptic efficacies that demonstrate different convergence locations. This characteristic is further demonstrated in Fig. 6b and 6c, where the values of synaptic efficacy in 100 independent training experiments were logged. It is demonstrated that each synaptic efficacy value was adjusted neighboring its initialization location, with minimal changes occurring from its starting value. As a result of this efficacy adjustment characteristic, a group of almost evenly distributed efficacy values exists between -0.4 and 0.4, even after hundreds of efficacy updates. This group exhibits a minor preference for an efficacy value of 0.1. In other words, the Tempotron did not converge the synaptic efficacies to the same result across different training processes. Instead, it updates the synaptic efficacies within a narrow range around their initialization values and is capable of finding a combination of efficacies that enables PSD in each independent experiment.

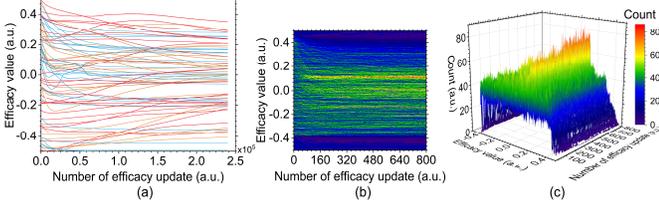

Fig. 6. Synaptic efficacy updates. (a) Synaptic efficacies in three independent instances of training. Each color represents the learning and updating of 25 synaptic efficacies during a training instance. (b) Two-dimensional and (c) three-dimensional heat maps of the distribution of efficacies in 100 independent training trials.

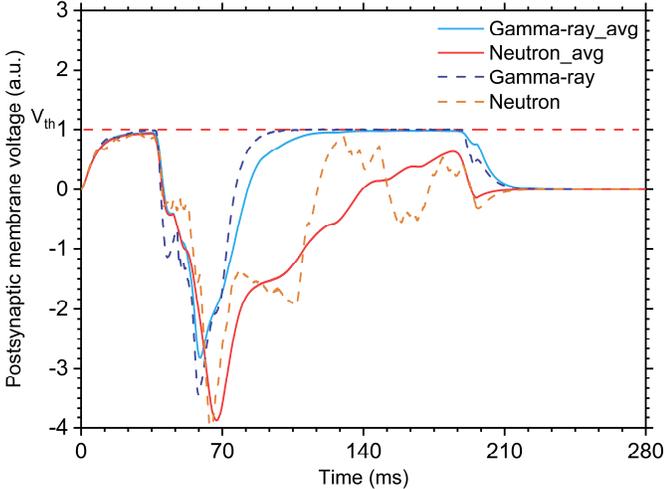

Fig. 7. Postsynaptic potentials. For neutron signals categorized as 0, the postsynaptic potential never exceeds the threshold. For gamma-ray signals categorized as 1, it exceeds the threshold once, generating a spike.

Fig. 7 demonstrates the postsynaptic potentials (PSPs) of the Tempotron. The Tempotron was trained with the following parameters: $\tau = 8.4$, $\tau_s = 2.1$, $dendrites_{num} = 25$, $epochs = 300$, $lr = [10^{-5}, 10^{-3}]$. The training employed momentum acceleration. When the Tempotron receives a gamma-ray signal, its PSPs exceed the threshold and produces a spike. Conversely, the PSPs do not reach the threshold if the input is a neutron event. The average PSPs of both types of particles are also presented, indicating that the PSPs of gamma-ray signal exceed the threshold while that of the neutron signal does not. Notably, different from the average PSPs presented in [24], it does not have a uniform distribution over time as the input spike patterns did not follow uniform distribution in the current study. However, the PSPs closely resemble the shape of the pulse signal depicted in Fig. 2a. This suggests that both synaptic plasticity and PSPs of the Tempotron are strongly correlated to input spatiotemporal spike patterns. The Tempotron learns from the encoded information, adjusting its synaptic efficacies to receive information inside signals, and expressing input signal features in its PSPs.

*F. Time-of-Flight*

All previous experiments employed ground truth labels derived using the PCNN method for training. This approach aligns with typical application scenarios for PSD methods, wherein neutron detection systems generally cannot provide ground truth particle labels. However, this practical strategy for preparing the training set is intrinsically doubtful in trustworthiness, thus leaving the performance of Tempotron potentially questionable. As a result, we conducted experiments using the Time-of-Flight (ToF) dataset, where the particle labels, verified via coincidence measurements, serve as a more reliable ground truth [43].

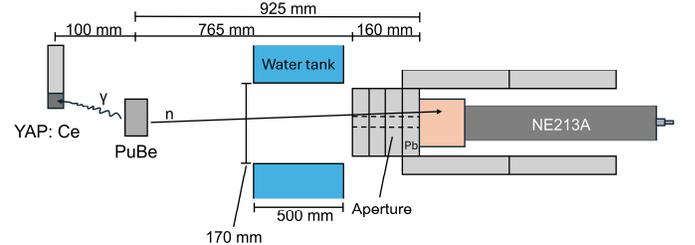

Fig. 8. Time-of-Flight experimental setups. YAP:Ce gamma-ray detectors and a NE213A neutron detector were used to perform coincidence measurements on the neutrons and energetic gamma-rays produced when alpha particles hit beryllium.

Fig. 8 presents a schematic representation of the experimental setup. The PuBe source is used for generating neutron and gamma-ray mixed field. Four YAP:Ce detectors are installed, each approximately 10 cm from the PuBe source, exhibiting minute out-of-plane position differences. These detectors measure low-energy cascade gamma rays from $^{234}$U and the energetic 4.44 MeV gamma-rays. These energetic gamma-rays are particularly interesting because they come from the nuclear reaction: $\alpha + ^{9}_{4}Be \rightarrow ^{12}_{6}C + n + \gamma$. The source and four gamma-ray detectors are housed within a water-encased shield cube, with a wall thickness of roughly 500 mm. A lead (Pb) enclosure is designed at the 170 mm diameter beam-port exit to house the liquid scintillator detector NE213A. The NE213A is positioned in the enclosure with the upstream face of the cell distanced 925 mm from the PuBe source center and at the same height as the source. The cylindrical symmetry axis of this detector was directly oriented towards the source center. A 10 mm × 10 mm aperture in the 160 mm wide face of the lead enclosure facilitated the measurement of source-related low energy cascade gamma rays and high energy 4.44 MeV gamma-rays. See more details in [44].

During the ToF measurement, both tagged neutron and prompt gamma-ray (or gamma flash) events were detected. A tagged neutron refers to a characteristic 4.44 MeV gamma ray detected in the YAP:Ce detector in coincidence to a fast neutron detected in the NE213A detector. On the other hand, prompt gamma-ray points to time-correlated gamma rays detected coincidentally in both detectors. Fig. 9a illustrates the recorded

TABLE II
DISCRIMINATION ACCURACY OF VARIOUS METHODOLOGIES

| Method | CC | ZC | CI | PGA | FGA | PCNN |
|---|---|---|---|---|---|---|
| Accuracy | 0.9381 | 0.5377 | 0.8160 | 0.4387 | 0.6736 | 0.9497 |
| Method | HQC-SCM | LG | KNN | MLP-I | MLP-II | Tempotron |
| Accuracy | 0.8952 | 0.8634 | 0.9624 | 0.8985 | 0.8994 | 0.9515 |
| Trainable Parameters | - - | - - | - - | 301 | 329,345 | 25 |

ToF spectrum. We deemed events within the (0.5 - 6) ns interval as the prompt gamma events, whereas those within the (28.1 - 60) ns interval were recognized as the tagged neutron events. All non-prompt gamma events were culled from the tagged neutron events via a charge integration threshold [43]. The derived neutron and gamma-ray spectrum is displayed in Fig. 9b. Preprocessing with a 1.5 MeV software threshold and removal of significantly corrupted signals, the ToF dataset comprises roughly 30,000 signals, each with its assigned ground truth label.

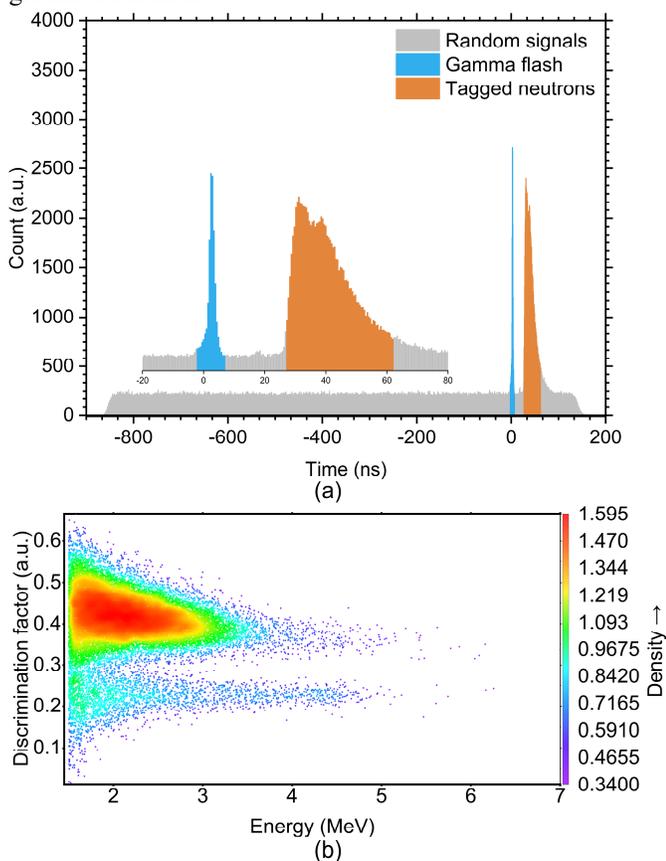

Fig. 9. Pulse signals in the Time-of-Flight experiment. (a) Time-of-Flight spectrum. (b) Energy with respect to the discrimination factor of charge integration method.

Approximately 6% of the ToF dataset, 2,000 pulse signals, were utilized for training set. Among the training set, 400 signals were reserved for validation, while 1,600 signals were used for training. The rest 94% of the ToF dataset composed testing set. The PSD results on the ToF dataset are shown in Table II, in which the Tempotron was trained with the following parameters: $\tau = 9$, $\tau_s = 2.25$, $dendrites_{num} = 25$, $epochs = 150$, $lr = [10^{-5}, 10^{-3}]$. The training employed both noise augmentation (with $\sigma_G = \sigma_j = p = 10^{-4}$) and momentum acceleration. Moreover, the accuracies are on the testing set.

As demonstrated in Table II, for these data and experimental setup, ZC, PGA and FGA display inadequate discrimination performance once more, securing accuracy rates below 80%. Conversely, other statistical PSD methods manifest accuracy spanning from 80% to 95%, with PCNN marking the highest accuracy at 94.97%. These findings align with those obtained from experiments conducted using the AmBe source in previous sections, further validating the robustness and applicability of employing PCNN as the ground truth labeler. Concerning the four machine learning methods, both KNN and Tempotron present superior performance with PSD accuracies exceeding 95%, whereas the two MLP methods hover around 90% in accuracy. Note that the discrimination accuracies on the training, testing, and validation sets are consistent across all machine learning methods, suggesting the absence of overfitting. When compared to results from the AmBe source experiment, the accuracy of MLP methods had a downfall in the ToF PuBe source experiment, while the Tempotron saw a rise. This shift originates from the lower energy threshold and more lenient corrupted signal exclusion utilized in the ToF dataset preparation. The ToF dataset signals span a broader energy range and are more susceptible to noise and signal corruption, which lends a more non-linear attribute to its discrimination task compared to that of the AmBe dataset. As a result, non-linear classifiers such as Tempotron and KNN outperformed the linear MLP-I classifier and the not sophisticated enough MLP-II classifier.

## IV. DISCUSSION

Previous experiments have demonstrated that the Tempotron is a highly effective classifier. In a few epochs and trainable parameters, it learns very quickly, achieving an accuracy of over 80%. It is highly resistant to noise in both signals and spike times. Additionally, the Tempotron is not reliant on efficacy value initialization and can realize its functionality using numerous combinations of efficacy values. Importantly, it can effectively adjust these efficacies to accurately extract information from input spatiotemporal spike patterns, leading to high classification accuracy.

The Tempotron model is inherently non-linear because of the spike timing-based input-output dynamics. This intrinsic non-linearity makes spike-based neural networks superior learners to perceptron-based neural networks [24]. Recent breakthroughs in large scale pre-trained transformers in both natural language processing [45, 46] and image processing [47, 48] have demonstrated the exceptional abilities of second-generation neural networks. However, the development of spike-based third-generation models is still in its early stages. By leveraging GPU implementation, third-generation neural networks such as the Tempotron can benefit from the fast-growing field of second-generation neural networks. Furthermore, spike-based neural networks such as the Tempotron can potentially be implemented biologically [49], which could greatly reduce the computational burden on digital computers.

Several limitations exist regarding the Tempotron and its application in PSD. Firstly, it is a binary classifier, capable of distinguishing between only two types of input patterns: those of neutron and gamma-ray pulse signals. Real-world radiation pulse signal data, however, is often corrupted by various sources, such as signal pile-up events. The Tempotron is not equipped to handle these significantly corrupted pulses without preliminary processing and their elimination from the dataset. Secondly, it does not benefit from large batch-size training; the reduction in Tempotron loss comes from averaging no more than 20 signals. This phenomenon arises from its learning paradigm, which differs from the established gradient-descent principle of deep learning models. When the synaptic efficacy adjustment is averaged over too many inputs, this update no longer represents the correct efficacy update direction. Finally, despite being a third-generation model, the Tempotron still requires several manually tuned hyperparameters, guided by the individual expertise of researchers, similar to second-generation deep learning models. Although the number of hyperparameters and trainable parameters is significantly lower than that of second-generation deep learning models, the necessity for multidisciplinary expertise poses challenges for the application of the Tempotron in PSD.

## V. Conclusion

In conclusion, this study details the implementation of the Tempotron, a third-generation neural network model, for discriminating pulse shapes in radiation detection. It encodes pulse signals into spike patterns using latency and Gaussian receptive fields. By utilizing multiple axon terminals of a presynaptic neuron, it encodes time-domain pulse signals into spatiotemporal spike patterns. The encoded spike patterns are then transmitted to the Tempotron neuron through synapses formed between the presynaptic axons and postsynaptic dendrites. The Tempotron neuron learns to discriminate between pulse signals of two particle types by adjusting its synaptic efficacies. The implementation of Tempotron utilizes GPU acceleration based on PyTorch in Python to achieve faster processing speed. This implementation is publicly available on GitHub at https://github.com/HaoranLiu507/TempotronGPU.

The experimental results indicate that the Tempotron model performs classification efficiently and effectively. Given its ability to process pulse signals directly and make decisions based on prior knowledge, it eliminates the need for feature extraction. The study also explored the impact of noise augmentation on the performance of Tempotron. The findings indicate that the appropriate level of noise intensity can boost generalization. Moreover, this study delved into the neural activity of Tempotron, elucidating its intrinsic data processing principles and guiding in its implementation. This study culminated in examining the reliability of Tempotron in PSD using the ToF methodology, which offered tagged neutrons and prompt gamma-rays as the ground truth for training. The experimental results reflected the robustness of Tempotron, particularly when dealing with datasets spanning a broad energy range and subject to moderate corruption.

However, utilizing the Tempotron in PSD is limited to processing not significantly corrupted pulse signals, small batch size, and the reliance on manual hyperparameter tuning. Nonetheless, due to its intrinsic non-linearity and biologically plausible gradient-based learning algorithm, the Tempotron remains a promising technology for dealing with PSD and other classification problems, including image and voice classification. Future research efforts should seek to investigate the use of the Tempotron for PSD on datasets with significantly corrupted signals, while developing techniques to reduce the reliance on hyperparameter tuning.

## Acknowledgment

The authors thank Nicholai Mauritzson and Kevin G Fissum from Division of Nuclear Physics, Lund University, for invaluable assistance regarding Time-of-Flight methodology. Furthermore, the authors express their appreciation to Feixiang Zhao, Xue Zuo, and Huibin Li from Chengdu University of Technology, for insightful discussions. Additionally, the authors extend their gratitude to Jialin Wu from School of Design Art, Changsha University of Science & Technology, for figure conceptional design and visualization. Finally, the authors extend their sincere appreciation to the anonymous peer reviewers for their valuable feedback and critiques that greatly enhance the quality of this research paper.